\documentclass{iau}
\usepackage{graphicx} 

\title[IAUS291.~~{\slshape Fermi}-LAT searches for $\gamma$-ray pulsars] 
{{\slshape Fermi}-LAT searches for $\gamma$-ray pulsars} 

\author[Saz Parkinson]  
{P.~M.~Saz Parkinson {\small for the {\slshape Fermi} LAT Collaboration}}

\affiliation{Santa Cruz Institute for Particle Physics,
  University of California, Santa Cruz, CA 95064 \\ email: {\tt pablo@scipp.ucsc.edu} \\[\affilskip]
}

\pubyear{2012}
\volume{291}  
\jname{\mbox{Neutron Stars and Pulsars: Challenges and Opportunities after 80 years}}
\editors{J. van Leeuwen, ed.} 
\begin{document}

\maketitle

\begin{abstract}
The Large Area Telescope (LAT) on the {\it Fermi} satellite is the
first $\gamma$-ray instrument to discover pulsars directly via their
$\gamma$-ray emission. Roughly one third of the 117
$\gamma$-ray pulsars detected by the LAT in its first three years were
discovered in blind searches of $\gamma$-ray data and most of these
are undetectable with current radio telescopes. I review some of the
key LAT results and highlight the specific challenges faced in
$\gamma$-ray (compared to radio) searches, most of which stem from
the long, sparse data sets and the broad,
energy-dependent point-spread function (PSF) of the LAT. I discuss some ongoing LAT searches for $\gamma$-ray millisecond pulsars
(MSPs) and $\gamma$-ray pulsars around the Galactic Center. Finally, I outline the prospects for future $\gamma$-ray pulsar
discoveries as the LAT enters its extended mission phase, including advantages of a possible modification of the LAT observing profile.
\keywords{pulsars: general, gamma rays: observations}
\end{abstract}


\firstsection 
\section{Introduction}

In the first four decades since their discovery, pulsars were almost 
the exclusive domain of radio (and to a lesser extent X-ray)
astronomy. Indeed, of the $\sim$2000 known pulsars\footnote{{\tt
    http://www.atnf.csiro.au/research/pulsar/psrcat}}, the majority were
discovered in radio. Since June 2008, however, with the launch of the
LAT, on the {\it Fermi} satellite, $\gamma$-rays have become a viable means of discovering (and studying)
pulsars. More important than the number of LAT-detected pulsars
(a small fraction of the overall population), the LAT sample is subject to
different biases than the radio sample. LAT pulsars are typically nearby ($\sim$few kpc) and energetic ($\dot{E}>10^{33}$ erg s$^{-1}$),
and a large fraction are {\it radio-quiet}. Furthermore,
$\gamma$-rays, unlike the radio beams, carry a significant fraction of the rotational energy of
pulsars, thus providing a powerful probe into these extreme objects. Finally, LAT pulsars provide a crucial input into our
understanding of the overall neutron star population of the Galaxy.

The LAT is a pair conversion telescope consisting of a tracker, a
calorimeter, and a segmented anti-coincidence
detector, along with a programmable trigger and a complex data
acquisition system. By incorporating the latest advances from particle
physics\footnote{Bill Atwood was awarded the 2012 Panofsky Prize in
  Experimental Particle Physics ``For his leading work on the design,
  construction, and use of the Large Area Telescope [...]''.}, including the use of silicon-strip detectors, the LAT has
achieved a giant leap in capabilities compared to its predecessor,
EGRET. The LAT extends to higher energies ($>$300 GeV vs $\sim$10
GeV), has a larger effective area and field of view, improved angular
and energy resolutions, and much lower deadtime (\cite[Atwood et al. 2009]{Atwood09}). In its first 4
years of operations, the LAT has collected $>$200 million ``source'' class events, compared to $\sim$1.5 million
photons collected by EGRET, in its 9-year lifetime.

Regarding pulsars, the progress made by the LAT has been
spectacular. The $\gamma$-ray pulsar population has grown from 7 firm
detections (plus a few candidates) at the time of the {\it Fermi}
launch (\cite[Thompson 2008]{Thompson08}), to 117 detections in three years of LAT
survey observations (The Second {\it Fermi}-LAT Catalog of
$\gamma$-ray Pulsars, {\it in preparation}). Beyond the jump in the number of pulsars, the increased
statistics enable detailed (e.g. phase-resolved) studies on individual
pulsars, previously out of reach. The LAT has
also opened up the unexplored 10--100 GeV window. EGRET
detected a mere handful of $>$10 GeV photons from the
brightest pulsars (\cite[Thompson 2005]{Thompson05}), whereas the LAT detects significant
emission from over two dozen pulsars in this energy range and even $>$25 GeV
events from the brightest ones (e.g.  Crab, Vela, Geminga, see
\cite[Saz Parkinson 2012]{Gamma2012}).

Of the 117 LAT detections, 61 pulsars were known prior to  {\it
  Fermi}. Pulsations were obtained by folding the $\gamma$-rays with a
radio (or X-ray) timing model. Of these 61, 20 are MSPs
(e.g. \cite[Abdo et al. 2009c]{MSP}), a hitherto unknown class of strong
$\gamma$-ray emitters. More surprisingly, a large number of MSPs ($>$40)
were discovered in radio searches of LAT
unassociated sources (\cite[Ray et al. 2012]{PSC}). Most (20 so far, e.g. \cite[Ransom et al. 2011]{Ransom11}) will
likely exhibit $\gamma$-ray pulsations\footnote{Given that they are
  LAT $\gamma$-ray sources selected precisely for their {\it pulsar-like} qualities.}, once enough data and/or a precise timing model is obtained (usually from
radio observations). Finally, 36 of the 117 $\gamma$-ray
pulsars were discovered directly in blind searches of LAT data
(\cite[Abdo et al. 2009b, Saz Parkinson et al. 2010, Saz Parkinson
2011, Pletsch et al. 2012a, Pletsch et al. 2012b]{BSPI,BSPII,Sardinia,Pletsch12,Pletsch12b}). The rest of this
paper discusses the challenges, results, and prospects of these searches.


\section{Blind searches for $\gamma$-ray pulsars (compared to radio)}

Two factors make $\gamma$-ray searches for pulsars particularly
challenging, compared to radio searches. The first involves the
scarcity of events. The LAT detects $\sim$1 $\gamma$-ray per day
from a typical (bright) pulsar. This means that LAT searches for pulsars span
months to years (see Figure 1). The second complication involves the
broad, energy-dependent PSF of the instrument (from $\sim5^\circ$ at 100 MeV to $\sim0.2^\circ$ at 100 GeV,
68\% containment, normal incidence). This results in significant
source confusion, especially in high background regions like the Galactic plane, where
the diffuse $\gamma$-ray emission makes it hard to resolve individual
sources. Essentially, it is impossible to select, with certainty,
events coming from a source, so maximum likelihood techniques must be
employed. Using these techniques, it is possible to determine the
probability that each event is coming from the source of interest and
improve the sensitivity of the search by assigning a weight to each event
equal to this probability \cite[(Kerr 2011)]{Kerr11}.

\begin{figure}[htbp]
\includegraphics[height=.25\textheight]{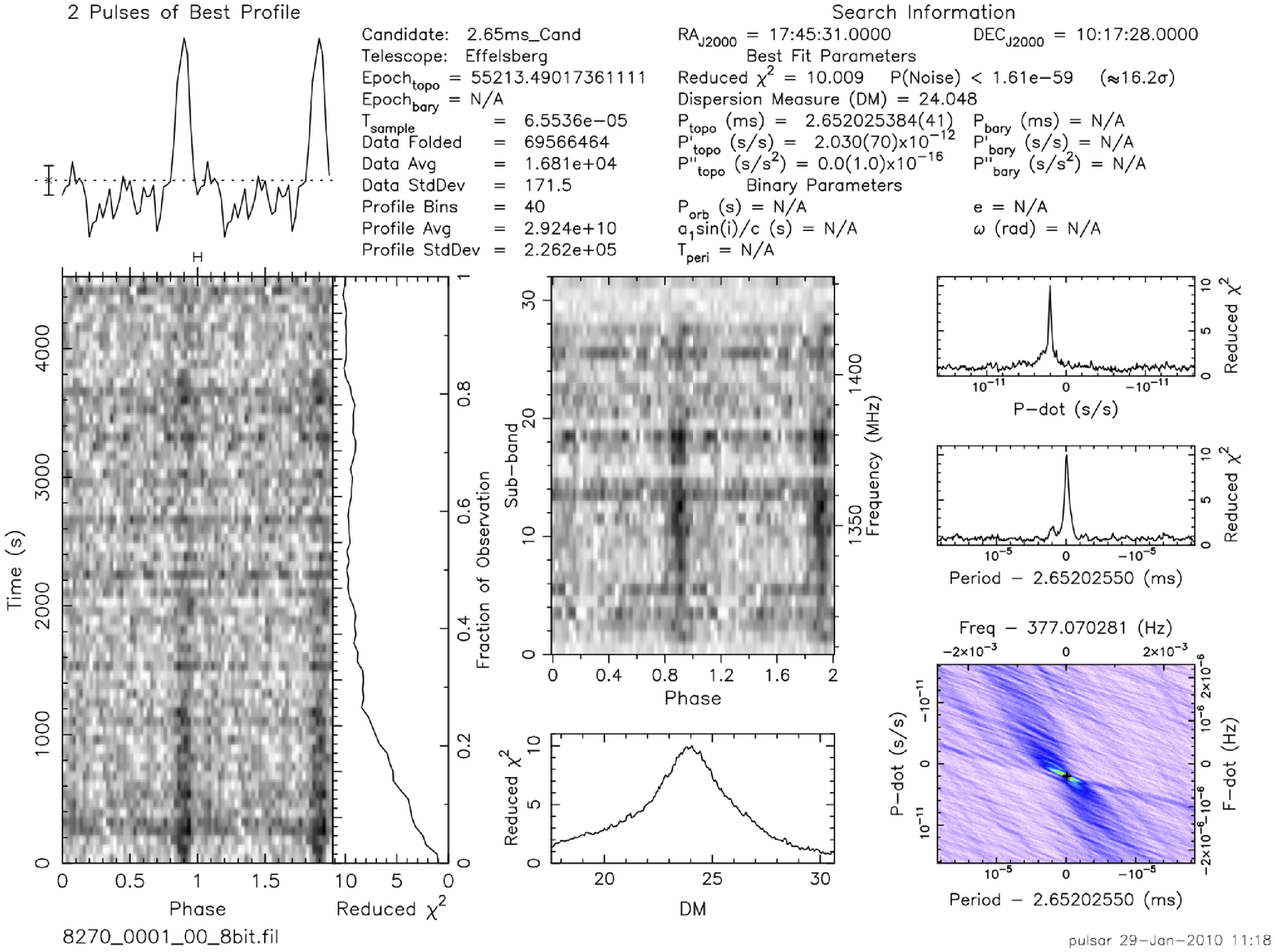}
\includegraphics[height=.25\textheight]{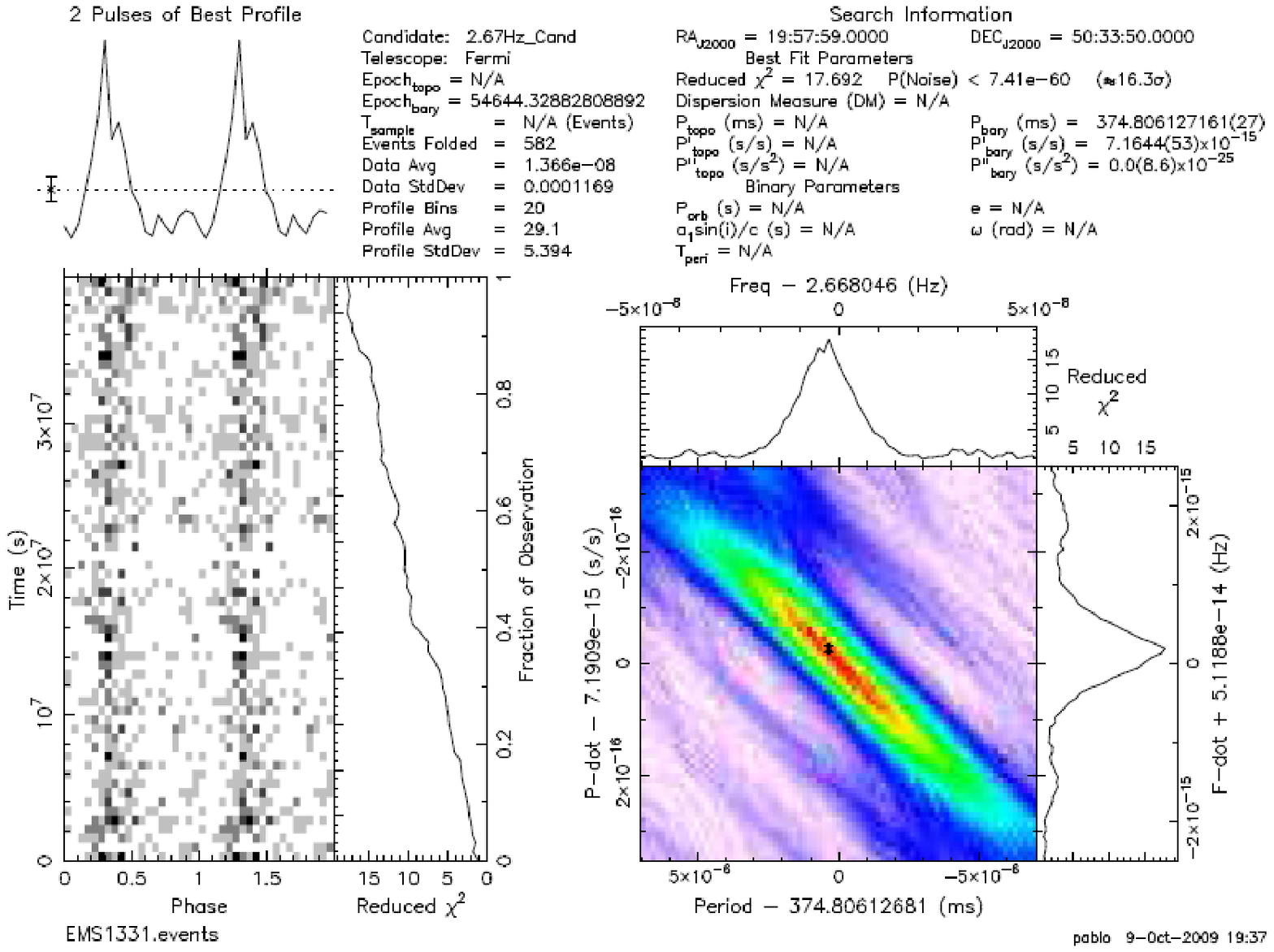}
\caption{ PRESTO (\cite[Ransom 2001]{Ransom01}) plots for two pulsars: one using radio data (left) and
  one using $\gamma$-ray data (right). {\bf Left} PSR J1745+1017, an MSP discovered in radio searches of {\it Fermi} unassociated sources
  with the Effelsberg telescope (Barr et al. 2012, MNRAS, {\it
    submitted}. Figure Credit: E. Barr). {\bf Right} PSR J1957+5033, a {\it radio-quiet} $\gamma$-ray pulsar discovered
  in a LAT blind search. Note that, although the pulse profiles and
  significances are similar, the $\gamma$-ray observation is $\sim$10,000 times
  longer, and the number of $\gamma$-ray events (582) is approximately
  one per day ($\sim$200,000 pulsar rotations!). On the other hand, unlike in radio, there is no need to
  scan in DM in a $\gamma$-ray search.}  
\label{fig6}
\end{figure}


Searches over long (sparse) data sets make standard FFT techniques
impractical. To tackle this problem, \cite[Atwood et al. (2006)]{Atwood06} developed the ``time-differencing
technique''  and applied it successfully to EGRET data
(\cite[Ziegler et al. 2008]{Ziegler08}). The core of the technique involves computing the
FFT of the differences between the times of events (up to a maximum,
sliding, time window), rather than using the original time series. Using a
time window of $\sim$weeks significantly alleviates the coherence requirements of the search.

\subsection{Young pulsars}

Within weeks of the launch of {\it Fermi}, the first {\it radio-quiet}
$\gamma$-ray pulsar was discovered in the supernova remnant CTA 1(\cite[Abdo et al. 2008]{CTA1}). This was
followed by many more discoveries (e.g. \cite[Abdo et al. 2009b, Saz Parkinson et
al. 2010]{BSPI,BSPII}). Many of these early
$\gamma$-ray pulsars discovered by the LAT are coincident with old
EGRET unidentified sources. Many, like the pulsar in CTA 1, are also
coincident with known supernova remnants or pulsar wind nebulae, and
were thus long suspected of hosting pulsars (e.g. PSR J1836+5925, J2021+4026). Radio follow-up
observations of the LAT-discovered pulsars showed that most of them are {\it radio-quiet} (or extremely
radio-faint), suggesting that the $\gamma$-ray emission originates far
from the neutron star surface, resulting in broad beams. It is now
clear that these radio-quiet $\gamma$-ray pulsars represent a significant fraction
of the neutron star population of the Galaxy (see Guillemot et al. (2012), in these proceedings, for a detailed
review of the radio observations of LAT $\gamma$-ray pulsars). All 36 pulsars found in LAT blind searches to date are young
($\tau<1\times10^7$ yr) energetic ($10^{33}$ erg s$^{-1} <\dot{E}< 10^{37}$ erg s$^{-1}$) pulsars with frequencies below
$\sim$20 Hz. These pulsars often exhibit timing irregularities, in
the form of {\it timing noise} and glitches. Since most are not detected in radio, the LAT is the only instrument
capable of timing them (\cite[Ray et al. 2011]{Ray11}). While it is
possible to time these noisy (or ``glitchy'') pulsars, a good timing
model often requires many frequency derivatives and other whitening
terms, making their {\it discovery} over long data spans extremely challenging. Searches for these pulsars may only be possible
over LAT observing periods lasting $\sim$months, rather than years (see Figure 2). 

\begin{figure}[b]
\begin{center}
\includegraphics[width=4.5in]{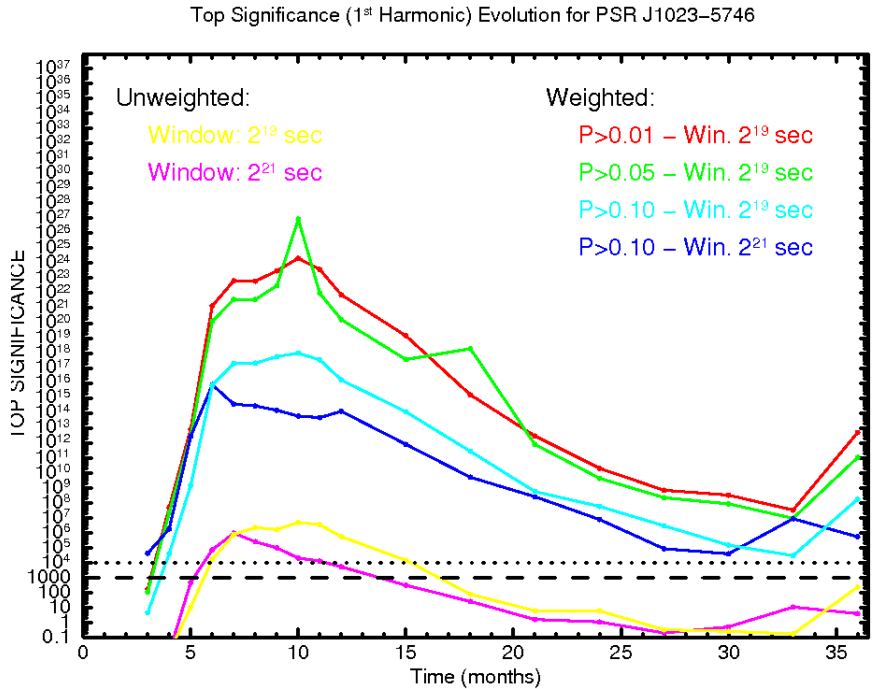}
\caption{Blind-search significance of PSR J1023-5745, as a function of the cumulative
  observing time. Initially, the significance increases with time,
  but after a period of 7-10 months, the significance peaks and then
  decreases with the addition of more data. 
}
\label{fig6}
\end{center}
\end{figure}

\subsection{Millisecond pulsars}

As mentioned above, the discovery of $\gamma$-ray MSPs (\cite[Abdo et
al. 2009c]{MSP}) was largely unexpected. Equally
unexpected was the large number of MSPs discovered by radio
telescopes searching LAT unassociated sources (\cite[Ray et al. 2012]{PSC}). This fact, combined
with the large fraction of {\it radio-quiet} young pulsars, raises the
question of whether large large numbers of {\it radio-quiet}
$\gamma$-ray MSPs remain to be discovered in blind searches of
LAT data. The answer, from examining the identified $\gamma$-ray
sources in the {\it almost} flux-limited {\it Fermi}-LAT Bright Source List (\cite[Abdo et al. 2009a]{BSL}),
appears to be no. Whereas at least two thirds of young $\gamma$-ray pulsars are {\it
  radio quiet}, at most one third of $\gamma$-ray MSPs are (\cite[Romani 2012]{Romani12}).

Blind searches for MSPs are vastly more complicated than searches for young pulsars. Firstly, a
majority ($\sim$80\%) of MSPs are in binary systems, and a full blind
search over unknown orbital parameters is out of reach given current
computer capabilities. Searches for isolated MSPs are possible, but
the high frequencies increase the memory and CPU
requirements. Positional uncertainties become more relevant with
increasing frequencies and the typical tolerance of a blind search
using several years of data is of order a fraction of an arc
second. This means that one must either perform a fine scan over the
LAT positions, or else identify a precise position of a plausible
counterpart, using multi-wavelength (e.g. X-ray) observations.
A number of LAT sources have been identified as promising pulsar
candidates, by virtue of their variability and spectral properties
(\cite[Ackermann et al. 2012]{logistic}). More recently, X-ray and optical studies
of some of the brightest of these have identified some strong
``black widow'' candidates: eclipsing binary MSPs with very compact, almost
circular orbits, where the pulsar is destroying its low-mass companion (\cite[Romani \& Shaw 2011, Romani
2012]{Romani11,Romani12}). These studies derive extremely
precise positions (to $\sim0.1^{\prime\prime}$) and stringent constraints on two
out of the three requisite orbital parameters (leaving only the projected semi-axis relatively
unconstrainted). Thus, for the first time, searches for such
binary MSPs using $\gamma$-ray data are possible. A deep LAT search
of 2FGL J2339.6+0532 (the most promising of these ``black
widow'' candidates) unfortunately produced no significant candidate,
but further efforts are in progress (see Belfiore (2012) for details).

\section{Searches for pulsars around the Galactic Center}
 
Understanding the $\gamma$-ray emission from the Galactic Center (GC)
region is both challenging and controversial. Claims that $\gamma$-ray emission from the GC may be related to dark
matter (e.g. \cite[Hooper \& Goodenough 2011, Weniger 2012]{HG11,Weniger12}) should be measured up against more conventional explanations, such as a possible origin from $\gamma$-ray pulsars.
As a site of massive star formation, it likely contains thousands of
pulsars, but their radio detection is hampered by the large
amount of interstellar scattering. Nevertheless, some pulsars have been discovered fairly close to
the GC, and predictions for the number of radio pulsars that could be
associated with the GC range in the thousands (\cite[Deneva et al. 2009]{Deneva09}). It
is possible that some of the undiscovered pulsars in the
region are {\it radio quiet}, like the majority of young pulsars found in
LAT blind searches (e.g. PSR J1732-3131, see Figure 3).

Blind searches for $\gamma$-ray pulsars around the GC are affected
by low fluxes (due to the large distance), and high levels of diffuse
emission. Indeed, most LAT pulsars are likely 
nearby ($\sim$few kpc). The LAT has, however, detected pulsations
from MSP J1823-3021A, in the globular cluster NGC 6624, at 8.4 kpc
(roughly the distance to the GC). It is possible to estimate how far blind searches might
be sensitive out to. From scaling arguments, considering that
the $\gamma$-ray flux of the Crab is several thousand times brighter than the faintest
$\gamma$-ray pulsar discovered in a blind search, we
conclude that it is possible to discover a {\it Crab-like}
$\gamma$-ray pulsar in a LAT blind search out to at least $\sim$15 kpc.

Perhaps the biggest complication in searching for faint pulsars around
the GC is the fact that such a pulsar may be young and {\it noisy} (or
``glitchy''). As described above, the loss of coherence of the signal limits the amount of data
that can be effectively searched (see Figure 2). Thus, regardless of
the number of years of LAT data available, it may only be possible to
search several months at a time. Thus, the sensitivity of searches in this region may only improve significantly
with improvements in reconstruction (e.g. Pass 8, see Baldini (2011),
{\it Fermi} Symposium), or a change in the observing mode. For the first
four years of its mission, {\it Fermi} has operated mostly in survey
mode. This has many advantages for most of the scientific goals of the mission, including pulsar searches (and
timing). Modifying the observing profile, however, could enhance the 
sensitivity to detection of faint signals (both from pulsars and/or dark
matter) from the GC. Figure 3 (right) shows the relative gain in
exposure (compared to survey mode) with a modified mode in which the
LAT points to a location slightly offset from the GC. Whenever the GC is
occulted, the LAT could go back into survey mode, thus maintaining
some level of exposure over the entire sky.

\begin{figure}[htbp]
\includegraphics[height=.19\textheight]{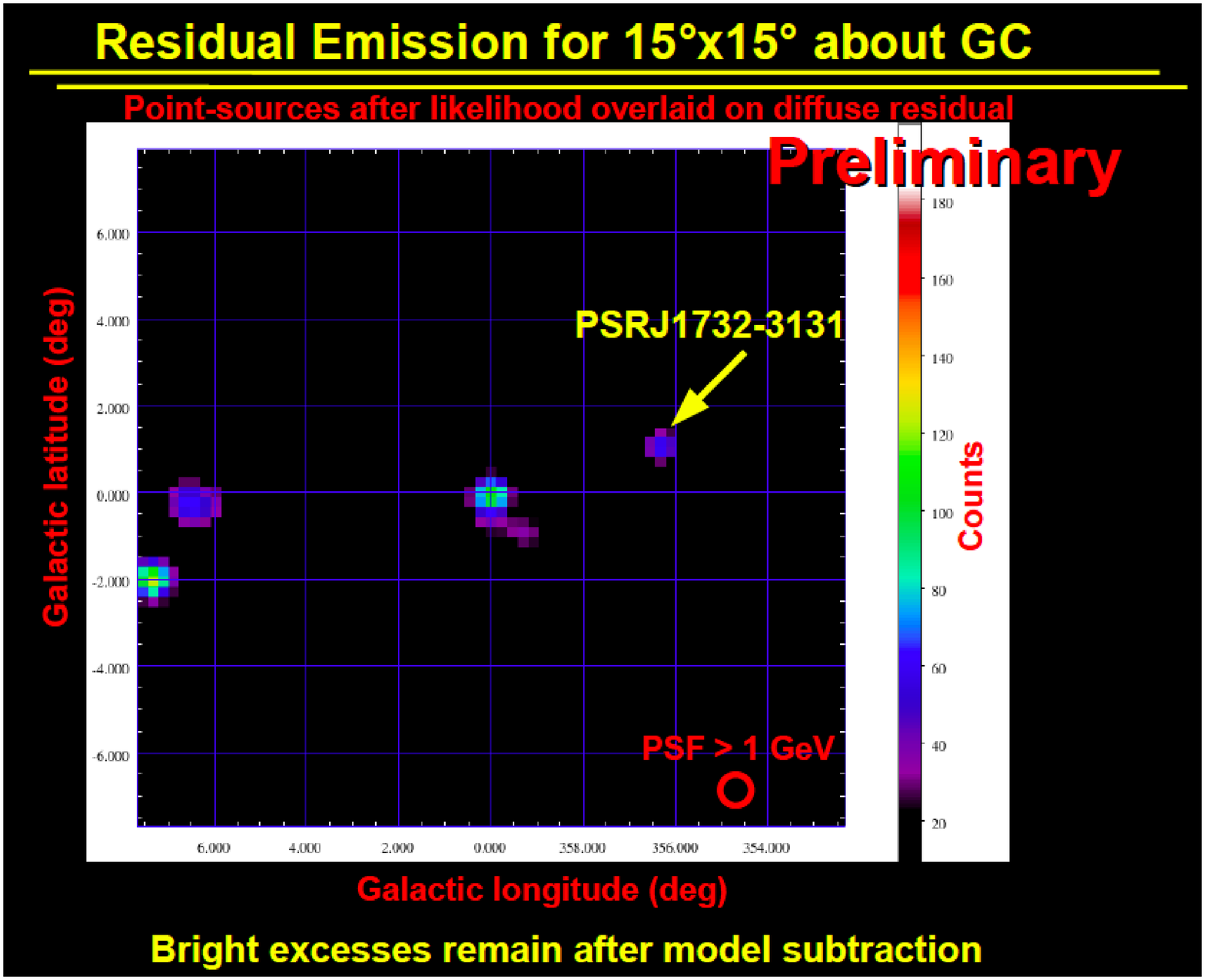}
\includegraphics[height=.20\textheight]{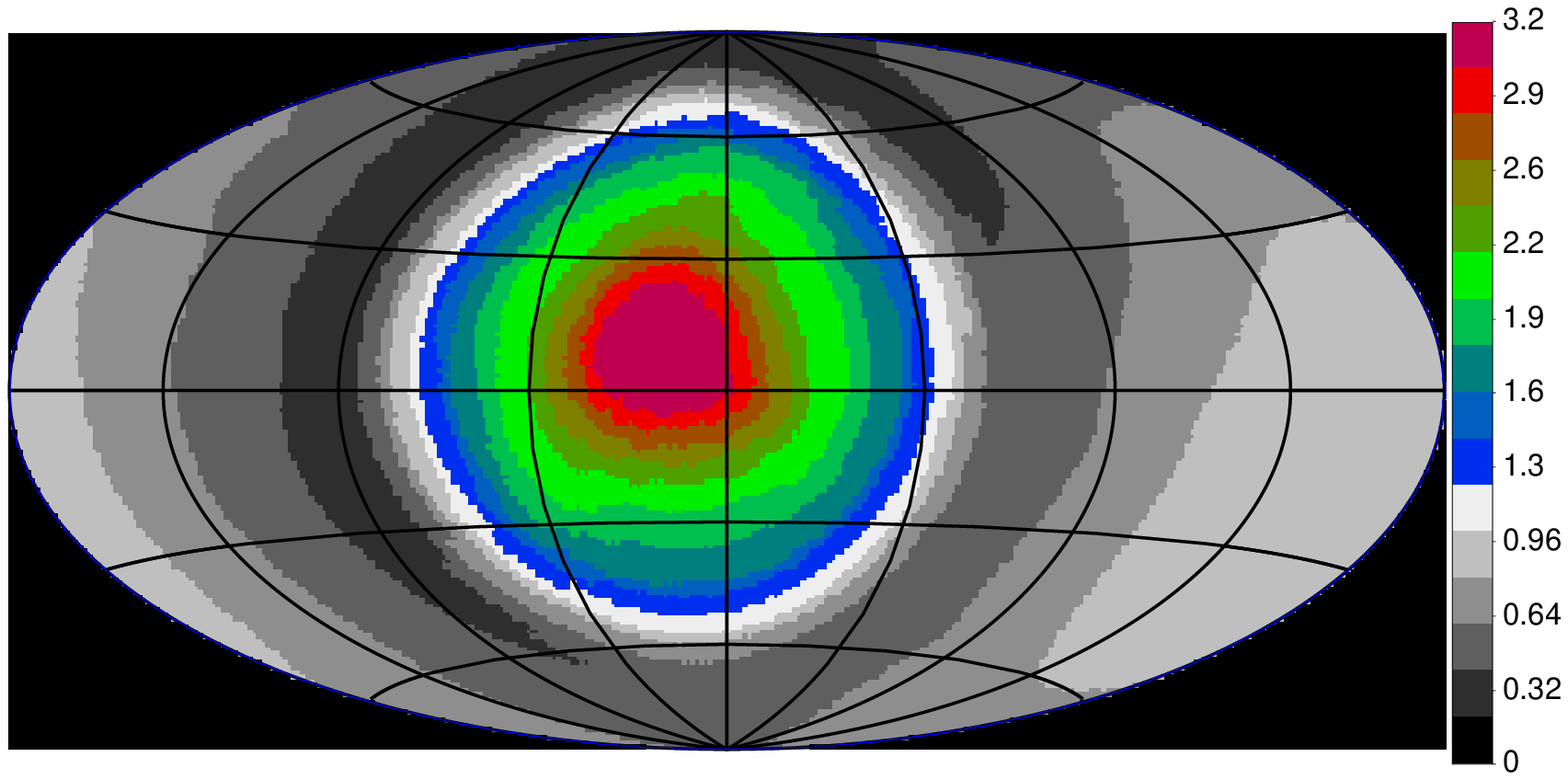}
\caption{{\bf Left} LAT view of the GC (Porter, 2011 {\it Fermi} Symposium). The bright source at the GC (2FGL J1745.6-2858) is curved and
  non-variable, like most LAT pulsars. {\bf Right}
  Relative exposure (compared to survey mode) with a LAT viewing
  strategy favoring the GC. The gain in exposure is a factor of
  $\sim$3, while the exposure for other parts of the sky decreases to
  roughly one third, in the worst case. }  
\label{fig6}
\end{figure}

\section{Conclusions and Prospects}
The NASA Senior Review recently recommended that {\it Fermi} operations
continue through 2016. In principle, {\it Fermi} could continue to
operate well beyond this date.\footnote{Unlike its predecessor,
  EGRET, the LAT has no consumables that limit its lifetime.} As the
LAT accumulates more data, it will detect $\gamma$-ray
pulsations from ever fainter {\it radio-loud} pulsars. As for blind
searches of $\gamma$-ray data, they too will continue to produce new
discoveries, although the increasing computational demands will
require additional resources (e.g. Einstein@Home\footnote{\tt{http://einstein.phys.uwm.edu}}) and
efficient computational techniques to exploit them.
Finally, multi-wavelength observations (radio, X-rays, and optical),
will continue to play a crucial role in LAT pulsar studies, especially
in facilitating the search for more exotic pulsar systems, such as
``black widow'' systems, young binary pulsars, and pulsars around the 
GC. In this last case, a modified observing profile
enhancing the exposure to this region (by a factor of $\sim$3) could
play a crucial role in future discoveries.

\section*{Acknowledgements}
The $Fermi$ LAT Collaboration acknowledges support from a number of
agencies and institutes for both development and the operation of the
LAT as well as scientific data analysis. These include NASA and DOE in
the United States, CEA/Irfu and IN2P3/CNRS in France, ASI and INFN in
Italy, MEXT, KEK, and JAXA in Japan, and the K.~A.~Wallenberg
Foundation, the Swedish Research Council and the National Space Board
in Sweden. Additional support from INAF in Italy and CNES in France
for science analysis during the operations phase is also gratefully
acknowledged. I acknowledge the support of the American Astronomical
Society and the National Science Foundation in the form of an
International Travel Grant, which enabled me to attend this conference.

\def \apjl {ApJL}
\def \apj {ApJ}
\def \apjs {ApJS}
\def \jcap {JCAP}

\end{document}